%
\catcode`@=11 
\font\titlefontB=cmssdc10 at 18pt
\font\sect=cmssdc10 at 12pt
\font\tmsm= zptmcm7t at 14pt
\font\matdop= msbm10 
%
 
%
\font\bfsymbtext=cmmib10 
\font\bfsymbscript=cmmib10 at 7pt
\font\bfsymbscriptscript=cmmib10 at 5 pt
\def\boldsymbol#1{{\mathchoice%
	{\hbox{$\displaystyle\textfont1=\bfsymbtext\scriptfont1=%
		\bfsymbscript\scriptscriptfont1=\bfsymbscriptscript #1$}}%
	{\hbox{$\textstyle\textfont1=\bfsymbtext\scriptfont1=%
		\bfsymbscript\scriptscriptfont1=\bfsymbscriptscript #1$}}%
	{\hbox{$\scriptstyle\scriptfont1=\bfsymbscript%
		\scriptscriptfont1=\bfsymbscriptscript #1$}}%
	{\hbox{$\scriptscriptstyle%
	\scriptscriptfont1=\bfsymbscriptscript {#1}$}}}}
%
\catcode`@=12 
\def\ny#1{\boldsymbol#1}
%
%
%
%

%
%
%
\def\downnormalfill{$\,\,\vrule depth4pt width0.4pt
\leaders\vrule depth 0pt height0.4pt\hfill\vrule depth4pt width0.4pt\,\,$}
\def\WT#1{\mathop{\vbox{\ialign{##\crcr\noalign{\kern3pt}
      \downnormalfill\crcr\noalign{\kern0.8pt\nointerlineskip}
      $\hfil\displaystyle{#1}\hfil$\crcr}}}\limits}

%
%
%
%
%
%
\magnification = \magstep1
\hsize = 16.9 truecm
\vsize = 21.6 truecm 
\parindent=0.6 truecm 
\parskip = 2pt
\normalbaselineskip = 12pt plus 0.2pt minus 0.1pt
\baselineskip = \normalbaselineskip 
\topskip= 17pt plus2pt      
\voffset= 0 truecm   
\hoffset= -0.01 truecm      
%
%
%
%
\def\tfract#1/#2{{\textstyle{\raise0.8pt\hbox{$\scriptstyle#1$}\over%
\hbox{\lower0.8pt\hbox{$\scriptstyle#2$}}}}}
\def\mezzo{\tfract 1/2 }
\def\quarto{\tfract 1/4 }

\def\sesto{\tfract 1/6 }
\def\rad{\tfract 1/{\sqrt 2} }

\def\radi2k{\tfract 1/{\sqrt {2k}} }
\def\der{\partial } 
\def\blk{\vrule height3.5pt width3.5pt depth0pt } 
\headline={\ifnum\pageno=0\line{ }\else {\vbox{\line{\hfil {\folio}}\line{\hrulefill}}}\fi}


\nopagenumbers
\count0=0

\null 

\line{\hfil IFUP-TH/2012-23}

\vskip 1.80 truecm 

\centerline {\titlefontB Representations for  creation  and annihilation operators}

\vskip 2.5 truecm 

\centerline {\tmsm Enore Guadagnini}

\vskip 1.0 truecm 

\centerline{\sl Dipartimento di Fisica, Universit\`a di Pisa,}
\centerline{\sl and INFN, Sezione di Pisa,}
\centerline{\sl Largo B. Pontecorvo 3, 56127 Pisa, Italy}

\vskip 3.4 truecm 

\centerline {\bf Abstract}

\medskip

\midinsert \narrower 
 A new representation ---which is similar to the Bargmann representation--- of the creation and annihilation operators is introduced,  in which  the operators act like  ``multiplication with" and like ``derivation with respect to" a single real variable. The Hilbert space structure of the corresponding states space is produced  and the relations  with the Schroedinger  representation are derived.  Possible connections of this new representation with the asymptotic wave functions of the  gauge-fixed quantum  Chern-Simons field theory and  (2+1) gravity are pointed out.  It is shown that the representation of the fields  operator algebra of the  Chern-Simons theory  in the Landau gauge is not a $*$-representation;   the consequences on the evolution of the states in the semiclassical approximation are discussed.

\endinsert 


\vfill\eject 

\noindent  {\sect  1. Introduction}  

\bigskip

\noindent The creation and annihilation operators that are associated with the one-particle  states play a fundamental role in  the perturbative approach to quantum field theories. The algebraic relations among these operators can be illustrated in the context of quantum mechanics. 
If $q$ and $p$ are canonical conjugated variables of a one-dimensional quantum mechanical system with $[q , p] = i $, one can introduce  the  annihilation $a_-$ and creation $a_+$ operators 
$$
a_- = \rad ( q + i p) \quad , \quad a_+ = \rad (q - i p) \; , 
\eqno(1)
$$
 satisfying the commutation relation 
 $$
 [ a_- , a_+ ] = 1 \; . 
 \eqno(2)
$$
In addition to the standard Schroedinger representation, the commutation relation (1.2)  admits the  Bargmann  representation [1] in which the operators 
$$
a_- = {\partial \over \partial z} \quad , \quad a_+ = z 
\eqno(3)
$$
act on the space ${\cal F }_A$ of the analytic functions ---in the complex plane---  of the complex variable  $z$ and  the inner product in ${\cal F}_A$ is given by 
$$
(f,g) = {1\over \pi } \int {\overline {f (z)}} \, g (z) \, e^{-  {\overline z} z } \; dz \, d{\overline z} \;  .  
\eqno(4)
$$
It is the purpose of the present paper to show that the defining relation (1.2)  also admits the representation in which the operators 
$$
a_- = {\partial \over \partial x} \quad , \quad a_+ = x 
\eqno(5)
$$
act on the space  of the  functions of the real variable $x$ that are the restriction on the real axis of analytic functions in the complex plane.  

The  spaces  of the wave functions for the ket and bra  state vectors  of the representation (1.5) will be described in detail and their connection with the conventional  Hilbert space ${\cal H}$ of the Schroedinger representation will be studied. The interplay between the Schroedinger and the new representations of the creation and annihilation operators  will be used to describe certain  representations of the  fields operator algebra ---which are not $*$-representations---  that  find a realization in the topological quantum  Chern-Simons field theory and (2+1) gravity  in the Landau gauge.   The wave functions associated with the asymptotic states of the gauge fixed theory will be described; it will be argued that, in the semiclassical approximation,  a nontrivial time evolution of the fields must always take place. 

\vskip 0.5 truecm

\noindent {\sect  2. The representation}

\vskip 0.4 truecm

\noindent   Let $\cal D$ be the algebra generated by the powers of $a_-$ and $a_+$ modulo the relation $[ a_- , a_+ ] = e $  where $e$ denotes the unit element of ${\cal D}$. As shown in Appendix A, the algebraic structure of $\cal D$  can be used to settle a relation between the  linear spaces of ket and bra vectors and certain elements of $\cal D$. 
 Given a ket vector $| e \rangle $ and a bra vector $\langle e | $ satisfying 
$$
a_- | e \rangle = 0 \qquad  , \qquad \langle e | a_+ = 0 \; , 
\eqno(6)
$$
and 
$$
\langle e | e \rangle =1 \; , 
\eqno(7)
$$
the space $V_-$ of ket vectors admits  
a basis given by the set of vectors $\{ | n \rangle =   (a_+^n / \sqrt {n!}  ) | e \rangle \} $ (with $n=0,1,2,...$). The vectors $\{ \langle m | =  \langle e | ( a_-^m / \sqrt { m!} ) \}$ form a basis in the space $V_+$  of bra vectors, and the pairing between $V_-$ and $V_+ $ is determined by   $\langle m | n \rangle = \delta_{m n}$. 
The algebra $\cal D$ is usually equipped with a $*$-involution given by 
$$
a_-^* = a_+ \qquad , \qquad a_+^* = a_- \; . 
\eqno(8)
$$

 The algebra generated by the  operators $a_- = \partial / \partial x $ and $a_+ = x $, acting on the functions of the real variable $x$,  is isomorphic with $\cal D$. The wave function $\psi_0 (x) $  which represents the ket vector  $ | e \rangle $ is essentially fixed by the condition  $a_- | e \rangle =0 $,  
$$
{\partial \over \partial x } \psi_0 (x) =  0 \; \Longrightarrow \; \psi_0 (x) = 1  \; . 
\eqno(9)
$$
Similarly, for the representative $\phi_0(x) $ of the bra vector $\langle e | $ satisfying $\langle e | a_+ =0 $,   one finds 
$$
 \phi_0 (x) x =  0 \; \Longrightarrow \; \phi_0 (x) = \delta (x)  \; . 
\eqno(10)
$$
The condition  $\langle e | e \rangle  = 1$ is indeed fulfilled 
$$
1 = \langle e | e \rangle =  \langle 0 | 0 \rangle \leftrightarrow  \int_{- \infty}^{+ \infty }  \phi_{0} (x) \psi_{0}(x) \, dx  = 
\int  \delta (x) 1 \, dx = 1   \; . 
\eqno(11)
$$
  The  wave functions that represent the ket vectors $ | n \rangle   \in V_- $ are $ | n \rangle \leftrightarrow \psi_n(x) = x^n / \sqrt {n!} $  (with $n=0,1,2,...$). The  functions $\{  \psi_n(x) \}$    constitute a basis for the linear space ${\cal G}_-$ of the wave functions associated with the ket state vectors.   As far as the bra vectors $ \langle m | \in V_+ $   are concerned,  the corresponding wave ``functions" (distributions) are $\langle m | \leftrightarrow  \phi_m(x) = (-1)^m \delta^{(m)}(x) / \sqrt {m!} $ (with $m=0,1,2,...$).  The elements $\{  \phi_m(x) \} $ compose a basis for the linear space ${\cal G}_+$ of the distributions that represent the bra state vectors. 
By construction, the elements $\psi_n (x)\in {\cal G}_-$  and the elements $ \phi_m (x)\in {\cal G}_+$ verify 
$$
\delta_{m n} = \langle m |  n \rangle \leftrightarrow  \int   \phi_m(x) \,  \psi_n(x) \, dx  = \delta_{m n }\; . 
\eqno(12)
$$

The conventional  Hilbert space structure of the  space $V_-$  is determined by the antilinear isomorphism $\sigma : V_- \rightarrow V_+ $ which is defined by  $ \sigma | n \rangle = \langle n | $. With the notation  $ \sigma  | u \rangle = \langle u_\sigma |  $, the inner product in $V_- $ of the element $|u \rangle $ and the element $| v \rangle $ is given by $ \langle u_\sigma | v \rangle $. Thus the canonical Hilbert space structure  of ${\cal G}_-$ is  induced by 
$$
 \sigma \; : \;   \psi_n (x) \rightarrow  \phi_n (x)  \; .  
 \eqno(13) 
$$
The decomposition of a generic wave function $\psi_A(x) \in {\cal G}_-$ with respect to the elements of the basis $\{ \psi_n(x) = x^n / \sqrt {n!} \}$ of ${\cal G}_-$ takes the form 
$$
\psi_A (x) = \sum_n {c_n \over \sqrt {n!}} x^n \; , 
\eqno(14)
$$
where $c_n$ are complex coefficients.  According to  the correspondence (13), the wave function $\sigma \, \psi_A (x) $ can be written as 
$$
\sigma \, \psi_A(x) = \sum_n {{\overline c_n} \over \sqrt {n!}} (-1)^n \delta^{(n)} (x) = {1\over 2 \pi } \int {\overline \psi }_A(ik) e^{ikx}  dk\; . 
\eqno(15)
$$
Consequently the inner product $\langle \psi_B | \psi_A \rangle $ of two wave functions $\psi_B(x) \in {\cal G}_-$ and $\psi_A(x)\in  {\cal G}_-$ ---where $\langle \psi_B |$ is represented by the wave function  $\sigma  \, \psi_B (x) $--- is given by 
$$
\langle \psi_B | \psi_A \rangle =   {1\over 2 \pi }\int   {\overline \psi_B}(ik) \, \psi_A(x) \, e^{ikx } \, dx  \, dk  \; . 
\eqno(16)
$$
Equation (16) takes the place of expression (4), which is valid for the inner product in the Bargmann representation. 
The wave  function $\psi_A(x)$ has finite norm when   $ | \psi_A |^2 = \langle \psi_A | \psi_A \rangle = \sum_n |c_n |^2 
$  is finite; in this case $\psi_A(x)\in {\cal G}_-$ is the restriction on the real axis $x \in \hbox{\matdop R} $ of a analytic function in the complex plane. 
This naturally fits the analytic structure [1] of the Bargmann representation. 

 In the new representation,  the  annihilation and creation operators $a_- = \partial / \partial x \, $ and $a_+ = x $  act on the linear space $\cal G_-$ equipped with the inner product (16);  these  operators   satisfy 
$$
  \left [ a_-\right ]^\dagger = a_+ \quad , \quad \left [ a_+ \right ]^\dagger = a_- \; . 
  \eqno(17)
$$ 
  Then if  $a_-^* = a_+$  ($a_+^* = a_- $), this representation  is a *-representation of the algebra $\cal D$. 
  
 The new representation  admits several variants; for instance, the choice   $a_- = x $ and $a_+= -\der / \der x $  constitutes  the mirror of the representation (5). For each couple of real parameters $x_0$ and $k_0$,    one can also introduce the nonstandard  representation 
$$ 
a_- =   {\der \over \der x} - i k_0 \quad , \quad  a_+ =   x  - x_0 \; , 
\eqno(18)
$$ 
in which the wave functions of the fundamental ket and bra vectors are 
$$
{\cal G}_- \ni \psi_0(x) =  e^{i k_0(x-x_0)} \quad , \quad {\cal G}_+ \ni  \phi_0(x) = \delta(x-x_0) \; . 
\eqno(19)
$$

A distinctive feature of the ket and bra wave functions of the representation (5) can be expressed by saying that, given a couple of  ``conjugated"  (in a generalized sense) variables like $\zeta$ and $\partial / \partial \zeta $,  the ket wave functions $\psi  \in {\cal G}_-$  depend for instance on the  values of the ``position" variable $\zeta $ exclusively, whereas the bra wave functions $\phi  \in {\cal G}_+$ only depend on the proper values of the corresponding ``momentum" $\partial / \partial \zeta $. 

\vskip 0.4 truecm 

\noindent {\sect 3. Connection with the Schroedinger representation \blk }
 In the  Schroedinger representation, where $q = x $ and $p = -i \der / \der x $, the operators
$$ 
a_-  = (x + \der / \der x)/ \sqrt 2 \quad , \quad a_+  = (x - \der / \der x ) / \sqrt 2  
\eqno(20)
$$
act on the space $\cal H $ of the wave functions of the real  variable $x$; the normalized  wave function $f_0 (x)$ which is associated with the vector $| e \rangle  $ is 
 $$
  f_0 (x)= (1 / \pi)^{1/4} \, \exp(- x^2 /2) \; . 
 \eqno(21)
 $$ 
The  representation of $a_-$ and $a_+$ of  Section~2 is physically equivalent to the Schroe\-din\-ger representation. The creation and annihilation operators  of the Schroedinger representation  can be obtained from the operators $a_- = \partial / \partial x $ and $a_+= x$  by means of a $\pi /4 $ rotation in the $(x , \der / \der x)$ plane  
 $$
{x + \der / \der x\over \sqrt 2}  = V \, (\der / \der x ) \, V^{-1} \quad , \quad {x - \der / \der x  \over \sqrt 2 }  = V \, x \, V^{-1} \; , 
\eqno(22)
$$
 where 
$$ 
 V =  \exp \left [ - {\pi \over 8} \left ( x^2 + {\der^2 \over \der x^2} \right ) \right ] \; .  
 \eqno(23)
 $$
Given the vector  $|A \rangle = \sum_{n=0} c_n \, | n \rangle \in V_-$, let $f_A(x) \in {\cal H}$ and $\psi_A(x)\in {\cal G}_-$   be the corresponding normalized wave functions in the Schroedinger representation and in the new representation  respectively. In agreement with equations (22) one finds 
$$
 \psi_A(x) = (2 \pi )^{1/4} \, V^{-1} f_A (x) \; .
 \eqno(24) 
$$
In order to prove equation (24) one only needs to consider the case in which  
$f_A(x)=f_0(x)$ and $\psi_A(x)=\psi_0(x)$. It is convenient to use the method presented by Bellazzini, Mintchev and Sorba in [2]. The operators $\{ Q = x^2 , R = \der^2 / \der x^2 ,  S = -2 - 4 x (\der / \der x ) \} $ generate the algebra with structure constants $[Q,R] = S $, $[S,Q]=-8 Q$ and $[S, R]= 8R$; thus in a neighborhood of the identity   one has 
$$
e^{\alpha (Q + R )} = e^{g(\alpha ) Q}\, 
e^{h(\alpha ) R}\, e^{k(\alpha ) S}\; , 
\eqno(25)
$$
where $\alpha $ is a real parameter and 
$g(\alpha )$, $h(\alpha )$ and $k(\alpha )$ are suitable functions of $\alpha$.  
The derivation of both sides of equation (25) with respect to  $\alpha $ determines   three differential equations for the functions 
$g(\alpha )$, $h(\alpha )$ and $k(\alpha )$
$$\eqalign { 
& 1 = g^\prime (\alpha ) + 4 \left ( h^\prime (\alpha ) - 8 k^\prime (\alpha ) h (\alpha ) \right ) g^2(\alpha ) + 8 k^\prime (\alpha ) g(\alpha ) \; , \cr
& 1 = h^\prime (\alpha ) - 8 k^\prime (\alpha ) h(\alpha ) \; , \cr
& 0 = g(\alpha ) \left ( h^\prime (\alpha ) - 8 k^\prime (\alpha ) h(\alpha)  \right ) + k^\prime (\alpha ) \; . \cr }
\eqno(26)
$$
Equations (26)  follow from the  structure constants  of the algebra generated by $Q$, $R$ and $S$ exclusively.   The integration of equations (26) produces the result 
$$\eqalign { 
&g(\alpha )= \mezzo \, {\rm {tg}} (2 \alpha ) \quad , \quad h(\alpha )= \quarto \sin (4 \alpha ) \; , \cr 
&k(\alpha )= \quarto \ln | \cos (2 \alpha ) | \quad , \quad (- \pi/4  < \alpha < \pi /4 )\; . \cr}
\eqno(27)  
 $$
When $\alpha = - \pi / 8 $, equation (25) gives 
$$
e^{-  (\pi / 8)  ( x^2 + \der^2 / \der x^2  )}  \cdot 1 = 2^{1/4} e^{- x^2 / 2 } \; . 
\eqno(28)
$$
By taking into account  equation (21) and the definition (23), one finds that equation (28) is equivalent to  $V \psi_0 (x) = (2 \pi)^{1/4} f_0 (x)$, and this concludes the proof of equation (24). 

The connection between the Schroedinger representation and the   representation of Section~2 is determined by  equation (24); the connections between the Schroedinger and the Bargmann representations have been derived in [1]. 

\vskip 0.5 truecm 

\noindent {\sect 4. Field theory perspectives }  

\vskip 0.4 truecm

\noindent The linear spaces  ${\cal G}_-$ and ${\cal G}_+$ of the wave functions for ket and bra vectors  ---presented in the previous sections--- can find applications not only in representation theory but also in the  solution of dynamical issues in quantum  systems.  The remaining part of this paper  is dedicated to this topic. The three dimensional  quantum field theory which is defined by the Chern-Simons lagrangian will be considered as an illustrative example. In field theories one has to do with an infinite number of degrees of freedom and consequently with an  infinite number of creation/annihilation  operators. The creation and annihilation operators of the new representation may  act by ``multiplication with"  and ``derivation with respect to"  the variables which are associated with the modes of  field configurations.

\vskip 0.4 truecm

\noindent {\sect 4.1. Vacuum rearrangement  \blk }   In the following exposition, the Fock space $\cal F$  canonically associated with the annihilation and creation  operators $a_-$ and $a_+$ will conventionally mean a linear space spanned by  $\{ | n \rangle = (a_+^n / \sqrt {n!}  ) | 0 \rangle \}    $ (with $n=0,1,2,...$) ---where the vector $| 0 \rangle \in {\cal F}$ satisfies $a_- | 0 \rangle =0 $---  equipped with the scalar product $\langle  m | n \rangle = \delta_{m n}$ so that,  with respect to the  the antilinear isomorphism $\sigma | n \rangle = \langle n | $, hermitian conjugation acts as $\left [ a_-\right ]^\dagger = a_+$ ($\left [ a_+ \right ]^\dagger = a_-$).   The action of the $*$-involution on the creation/annihilation  operators is not fixed a priori. Examples    will be presented in which $a^*_- \not= a_+$ ($a_+^* \not= a_-$); in this case, the representation of  $a_-$ and $a_+$ defined in $\cal F$  does not give a $*$-representation of the operator algebra. Consequently, the vacuum functional ---which is given by   the mean value of the operators with respect to $| 0 \rangle \in {\cal F}$---   may not necessarily be positive with respect to the $*$-involution but it will  always be positive with respect to the hermitian conjugation.  
 
In order to illustrate how representations of the operator algebra that are not $*$-re\-pre\-sen\-ta\-tions may take place, let us assume that,  in a given model, $( a_+ , a_- ) $ and $( b_+ , b_- )$  are two couples of  creation/annihilation operators, with the $a$-operators commuting with the $b$-operators.  The linear combinations 
$$\eqalign {
&\alpha_- = \mezzo \left ( a_- - a_+ + b_- - b_+ \right )  \; , \; \alpha_+ = \mezzo \left ( a_- + a_+ + b_- + b_+ \right ) \cr
&\beta_- = \mezzo \left ( a_- - a_+ - b_- + b_+ \right )  \; , \; \beta_+ = \mezzo \left ( a_- + a_+ - b_- - b_+ \right ) \cr } 
\eqno(29)
$$
maintain the commutation relations and define two couples $( \alpha_+ , \alpha_- ) $ and $( \beta_+ , \beta_- )$ of creation/annihilation operators.  When $( a_- , a_+ ) $ and $( b_- , b_+ )$ are expressed in the Schroedinger representation (20),
 $( \alpha_- , \alpha_+ ) $ and $( \beta_- , \beta_+ )$ are creation/annihilation operators in the representation (5).  
If the star involution acts as $a_-^* = a_+ $ and $b_-^* = b_+$, then one finds $\alpha_-^* = - \alpha_-$, $\alpha_+^* =  \alpha_+$ and similarly  $\beta_-^* = - \beta_-$, $\beta_+^* =  \beta_+$. 

The Fock space ${\cal F}_1$ which is canonically associated with $( a_- , a_+ ) $ and $( b_- , b_+ )$ furnishes the operator algebra with a $*$-representation. 
Instead, the representation of the operator algebra in the Fock space ${\cal F}_2$,  which is canonically associated with $( \alpha_- , \alpha_+ ) $ and $( \beta_- , \beta_+ )$, is not a $*$-representation. 
The Fock space ${\cal F}_1$  differs from  ${\cal F}_2$, indeed relations (29) define a generalized Bogoliubov transformation [3] and the change ${\cal F}_1\rightarrow {\cal F}_2$ corresponds to a rearrangement of the vacuum structure.  Bogoliubov transformations usually do respect the adjoint whereas transformations (29) do not, for this reason equations (29) are called  generalized Bogoliubov transformations.  

In quantum field theories,  the space of ket vectors and the space of bra vectors are usually related with the asymptotic behavior of the solutions of the equations of motion for the classical field variables and may also correspond to the asymptotic states spaces of incoming and outcoming particles in  scattering processes.  The ket and bra wave functions for the so-called one-particle states  determine the quantum numbers of the creation/annihilation  operators which enter the mode decomposition of the field operators in the interaction picture; therefore a change of the asymptotic Fock spaces ---as the change ${\cal F}_1\rightarrow {\cal F}_2$ mentioned above--- in general modifies the expectation values of the observables and it may modify the  interpretation of the physical content of the model. The vacuum rearrangement which is associated with the introduction of the new creation/annihilation operators (29) and the change of the states spaces ${\cal F}_1 \rightarrow {\cal F}_2$ is somehow similar (but it is not equal) to the introduction of the Dirac sea for fermions;   the vacuum settlements  discussed in the following sections concern bosonic states.

\vskip 0.4 truecm

\noindent {\sect 4.2. Field operators  in Chern-Simons theory \blk } One encounters the representation (5) ---for an infinite number of $a_-$ and $a_+$--- in the canonical quantization in $\hbox{\matdop R}^3$  of the Chern-Simons theory in the Landau gauge [4], when a Fock space representation [5] of the field operators  ---in the interaction picture---  is displayed.  The lagrangian density of the model  (defined in $\hbox{\matdop R}^3$)  is 
$$\eqalign { 
{\cal L} = & \mezzo \varepsilon^{\mu \nu \rho } A^a_\mu \partial_\nu A^a_\rho + \der_\mu B^a A_\nu^a \, \eta^{\mu \nu }  + \der_\mu {\overline c}^a \der_\nu c^a\,  \eta^{\mu \nu} \cr 
 &   -  \sesto  g \varepsilon^{\mu \nu \rho } \,  f^{abc }\,  A^a_\mu A_\nu^b A^c_\rho -  g f^{abd} \, \der_\mu {\overline c}^a A^b_\nu c^d\,  \eta^{\mu \nu} \; , \cr } 
\eqno{(30)}
$$
where the vector fields $A_\mu^a(x)$ represent the gauge connection, $B^a(x)$ is called the auxiliary field, $\{ f^{abc} \}$ are the structure constants of the gauge group $SU(N)$, $g$ is the  coupling constant and $\eta_{\mu \nu}$ denotes the background metric which enters the gauge fixing terms exclusively.  The ghost $c^a(x)$ and antighost ${\overline c}^a (x)$ operators admit the standard  mode decomposition in terms of creation and annihilation operators like in ordinary (non-topological) gauge theories. 
Let us  consider the field operators $A^a_\mu(x)$ and $B^a(x)$; assuming that the third component $x^3=t$ of the cartesian coordinates $x^\mu = (x^1, x^2 , x^3 )$ of  $\hbox{\matdop R}^3$ represents the ``time" component, one can write 
$$ \eqalign { 
 A^a_1 (x) &=  \int {d^2k \over 2 \pi} \rad \bigg \{ \cos \theta_{\ny {k}} \left [ u_-^a(\boldsymbol {k}, t) e^{i\boldsymbol {k x}} + u_+^{a} ( \boldsymbol {k}, t) e^{-i \boldsymbol {k  x}} \right ]  \cr 
 & \qquad \qquad ~ - \sin \theta_{\ny {k}} \left [ -i u_-^a(\boldsymbol {k}, t) e^{i\boldsymbol {k  x}} + i u_+^{a  } ( \boldsymbol {k}, t) e^{-i \boldsymbol {k  x}} \right ] \bigg \} \; , \cr}
 \eqno(31)
 $$
$$\eqalign { 
 A^a_2 (x) &=  \int {d^2k \over 2 \pi} \rad \bigg \{ \cos \theta_{\ny {k}} \left [ - i u_-^a(\ny {k}, t) e^{i\boldsymbol {k x}} + i u_+^{a  } ( \ny {k}, t) e^{-i \boldsymbol {k x}} \right ]  \cr 
 & \qquad  \qquad ~~~  + \sin \theta_{\ny {k}} \left [ u_-^a(\ny {k}, t) e^{i\boldsymbol {k  x}} +  u_+^{a}( \ny {k}, t) e^{-i \boldsymbol {k  x}} \right ] \bigg \} \; , \cr}
 \eqno(32)
 $$
$$ 
 A^a_3 (x) = \int {d^2k \over 2 \pi} \rad \bigg \{  - i v_-^a(\ny {k}, t) e^{i\boldsymbol {k  x}} + i v_+^{a } ( \ny {k}, t) e^{-i \boldsymbol {k  x}}  \bigg \} \; , 
 \eqno(33)
 $$
 $$
 B^a (x) = \int {d^2k \over 2 \pi} \rad \bigg \{   v_-^a(\ny {k}, t) e^{i\boldsymbol {k  x}} + v_+^{a}( \ny {k}, t) e^{-i \boldsymbol {k  x}}  \bigg \} \; , 
 \eqno(34)
 $$
where  $e^{i\boldsymbol {k  x}} = \exp [ i (k_1 x^1 + k_2 x^2)] $ and  $ \cos \theta_{\ny {k}} = k_1 / k\, $,  $\sin \theta_{\ny {k}} = k_2 / k  \,$,  $  k = \sqrt {k_1^2 + k_2^2}\, $. 
In the mode decomposition (31)-(34)  one may require that the $*$-involution acts as  
$$
\left [ u_-^a(\ny {k}, t) \right ]^* = u_+^a(\ny {k}, t) \quad , \quad \left [ v_-^a(\ny {k}, t) \right ]^* = v_+^a(\ny {k}, t) \; ,  
\eqno(35)
$$
so that the field operators that represent the real  vector  fields $A^a_\mu(x)$ and the real auxiliary field   $B^a(x)$ verify  
$$
\left [ A^a_\mu (x) \right ]^* = A^a_\mu(x) \quad , \quad  \left [ B^a (x) \right ]^* = B^a (x) \; . 
\eqno(36)
$$ 
Suppose now that the gauge-fixing metric has euclidean signature and that, in order to simplify matters,  $\eta_{\mu \nu}$  coincides with the  $3\times 3$ identity matrix,  $ \eta_{\mu \nu} dx^\mu d x^\nu =  (dx^1)^2 + (dx^2)^2 + (dt)^2 $.  The time evolution of $\{ u_{\pm}^a(\ny {k}, t)\} $ and $\{ v_{\pm}^a(\ny {k}, t)\}$  is uniquely determined by the free equations of motion 
$$
\varepsilon^{\mu \nu \rho} \der_\nu  A^a_\rho (x)  + \der^\mu B^a (x) = 0 \quad , \quad \der^\mu A^a_\mu (x) = 0  \; ;  
\eqno(37)
$$
one finds 
$$\eqalign { 
u_-^a (\ny {k} , t ) &= u_-^a (\ny {k} ) \, {\rm Ch} (k t)  - v_+^{a  } (- \ny {k}  ) \, {\rm Sh} ( k t ) \; , \;  
 v_-^a (\ny {k} , t ) = v_-^a (\ny {k}  ) \, {\rm Ch} (k t)  - u_+^{a  } (- \ny {k}  ) \, {\rm Sh} ( k t ) \; , \cr  
  u_+^{a} (\ny {k} , t ) &= u_+^{a} (\ny {k}  ) \, {\rm Ch} (k t)  - v_-^a (- \ny {k}  ) \, {\rm Sh} ( k t ) \; , \; 
 v_+^{a} (\ny {k} , t ) = v_+^{a}(\ny {k}  ) \, {\rm Ch} (k t)  - u_-^a (- \ny {k}  ) \, {\rm Sh} ( k t )  \; . \cr }
 \eqno(38)
$$
The couples of conjugated field variables  are determined by the variations of the action  $S$ 
$$
{\delta S \over \delta (\partial_t A^a_1(x))} = A^a_2(x) \quad , \quad 
{\delta S \over \delta (\partial_t B^a(x))} = A^a_3(x) \; ; 
\eqno(39)
$$
the  rules of the canonical quantization give $$
[  A^a_1(\boldsymbol {x} , t) ,  A^b_2(\boldsymbol {y} , t)] = i \delta^{ab} \, \delta (\boldsymbol {x} - \boldsymbol {y}  ) = [  B^a (\boldsymbol {x} , t) ,  A^b_3(\boldsymbol {y} , t)] \; ,   
\eqno(40) 
$$
and all the remaining equal-time commutators must vanish. In agreement with the canonical commutation relations (40), one has 
$$\eqalign { 
&\left [ u_-^a(\ny {k } ) , u_+^{b}( \ny {p}  ) \right ] = \delta^{ab} \delta (\ny {k} -  \ny {p}  ) = \left  [ v_-^a(\ny {k} ) , v_+^{b} ( \ny {p } ) \right ]  \; , \cr  
&\left [ u_-^a(\ny {k } ) , v_-^b ( \ny {p } ) \right ] =
\left [ u_-^a(\ny {k } ) , v_+^{b}( \ny {p } )\right ] =  0 \; , \cr 
&\left [ u_+^{a} (\ny {k } ) , v_-^b ( \ny {p } ) \right  ] = \left [ u_+^{a} (\ny {k } ) , v_+^{b} ( \ny {p } ) \right  ] =0  \; ,     \cr}
\eqno(41)
$$
which show that $ u_{\pm }^a(\ny {k } ) $ and $v_{\pm}^a(\ny {k} )$ represent creation/annihilation operators, with $*$-in\-vo\-lu\-tion   $\left [ u_-^a(\ny {k}) \right ]^* = u_+^a(\ny {k}) $ and $\left [ v_-^a(\ny {k}) \right ]^* = v_+^a(\ny {k}) $.

Let ${\cal F}_1$ be the Fock space which is canonically associated with the operators $ u_{\pm }^a(\ny {k } ) $ and $v_{\pm}^a(\ny {k} )$; the representation of the operator algebra on ${\cal F}_1$ is a $*$-representation. The ``vacuum" vector $| \omega \rangle \in {\cal F}_1$ fulfills $u_{-}^a(\ny {k } ) | \omega \rangle = 0 $ and $v_{-}^a(\ny {k } ) | \omega \rangle = 0 $. 
The mean values  of the product of the field operators (31)-(34) computed with respect to $| \omega \rangle $ are not invariant under time translations. Moreover  the  two-point functions, for instance, which are computed with respect to $| \omega \rangle $ (and which are not vanishing)  are not well defined since they are formally infinite.  

Let us consider then the new creation/annihilation operators $\alpha_{\pm} (\ny k)$ and $\beta_{\pm }(\ny k)$ which are defined by means of the relations 
$$\eqalign { 
& \alpha_-^a (\ny {k } ) =  \mezzo \left [ u_-^a(\ny {k } ) + v_+^{a} (- \ny {k } )  + v_-^a(\ny {k } ) + u_+^{a} (- \ny {k } )   \right ] \; , \cr
& \alpha_+^a (\ny {k } ) =  \mezzo \left [ u_+^{a} (\ny {k } ) - v_-^{a} (- \ny {k } )  + v_+^{a}(\ny {k } ) - u_-^{a} (- \ny {k } )  \right ] \; , \cr
& \beta_-^a (\ny {k } ) = \mezzo \left [ u_-^a(\ny {k } ) + v_+^{a} (- \ny {k } )  - v_-^a(\ny {k } ) - u_+^{a} (- \ny {k } )  \right ] \; , 
\cr 
& \beta_+^a (\ny {k } ) = \mezzo \left [ u_+^{a} (\ny {k } ) - v_-^{a} (- \ny {k } )  - v_+^{a}(\ny {k } ) + u_-^{a} (- \ny {k } ) \right ]  \; . 
\cr}
\eqno(42)
  $$
 The nonvanishing commutators are $\left [ \alpha_-^a(\ny {k } ) , \alpha_+^{b}( \ny {p}  ) \right ] = \delta^{ab} \delta (\ny {k} -  \ny {p}  ) = \left  [ \beta_-^a(\ny {k} ) , \beta_+^{b} ( \ny {p } ) \right ] $. 
  The $*$-involution takes the form  
$$\eqalign { 
&\left [ \alpha^a_- (\ny {k}) \right ]^*  =   \alpha^a_-(- \ny{k})  \quad , \quad 
\left [ \alpha^a_+ (\ny {k}) \right ]^* =  - \alpha^a_+(- \ny{k}) \; , \cr 
&\left [ \beta^a_- (\ny {k}) \right ]^* = - \beta^a_-(- \ny{k})  \quad , \quad  \left [ \beta^a_+ (\ny {k}) \right ]^*  =  \beta^a_+(- \ny{k}) \; . \cr } 
\eqno(43)      
$$
Let ${\cal F}_2$ be the Fock space which is canonically associated with $\alpha_{\pm} (\ny k)$ and $\beta_{\pm }(\ny k)$; the representation of the operator algebra on 
${\cal F}_2$ is not a $*$-representation. 
The   ``vacuum" vector $|0 \rangle \in {\cal F}_2 $ is defined by $\alpha_{-} (\ny k) | 0 \rangle =0 $ and  $\beta_{-} (\ny k) | 0 \rangle =0 $ and describes a stationary state.   The expectation values of the field operators computed with respect to $|0 \rangle \in {\cal F}_2$ are invariant  under time translations; in particular, the vacuum  expectation values of the time-ordered product of two field operators (31)-(34) coincide [5] with the components of the Feynman propagator in the Landau gauge which can also be  obtained [4],  for instance, by means of the path-integral quantization. This implies that the  expectation values of the Wilson line operators that are computed ---in perturbation theory--- with respect to $| 0 \rangle \in {\cal F}_2$  correspond precisely  to the link invariants [6,7,8] of the Chern-Simons theory in $\hbox{\matdop R}^3$.  

Conditions $\alpha_-^a (\ny {k } )| 0 \rangle =0$ and $\beta_-^a (\ny {k } )| 0 \rangle =0 $ (for any $\ny k $) are equivalent to  the conditions $\left [ u_-^a(\ny {k } ) + v^{a}_+ (- \ny {k } )  \right ] | 0 \rangle =0$ and $\left [ u^{a}_+ (\ny {k } ) + v^{a}_- (- \ny {k } )  \right ] | 0 \rangle =0$. Let the annihilation and creation operators $u^a_-(\ny {k } ) $ and $ u^{a}_+ (\ny {k } )$ be represented in the Schroedinger representation by  $ ( \zeta^a_{\ny k} +  \der / \der \zeta^a_{\ny k} ) / \sqrt 2$ and $ ( \zeta^a_{\ny k} -  \der / \der \zeta^a_{\ny k} ) / \sqrt 2$ for a certain real variable $\zeta^a_{\ny k}$; similarly, let $v^a_-(- \ny {k } ) $ and $ v^{a}_+ (- \ny {k } )$ be represented by $ ( \eta^a_{\ny k} +  \der / \der \eta^a_{\ny k} ) / \sqrt 2$ and $ ( \eta^a_{\ny k} -  \der / \der \eta^a_{\ny k} ) / \sqrt 2$. Then, the vector $| 0 \rangle $ satisfies 
$$
\left ( \zeta^a_{\ny k} + \eta^a_{\ny k} \right )  | 0 \rangle = 0 \quad , \quad \left ( \der / \der  \zeta^a_{\ny k} - \der / \der \eta^a_{\ny k} \right )  | 0 \rangle = 0 \; . 
\eqno(44)
$$
Similarly, conditions $\langle 0 | \alpha_+^a (\ny {k } ) =0 $ and $\langle 0 | \beta_+^a (\ny {k } ) =0  $ imply 
$$ 
\langle 0 |  \left ( \zeta^a_{\ny k} - \eta^a_{\ny k} \right )  = 0 \quad , \quad \langle 0 | 
\left ( \der / \der  \zeta^a_{\ny k} + \der / \der \eta^a_{\ny k} \right ) = 0 \; . 
\eqno(45)
$$
Equations (44) and (45) show that, as far as each variable  $\zeta^a_{\ny k} + \eta^a_{\ny k}$ is concerned,  the wave function associated with $ | 0 \rangle $ is an eigenfunction of the ``position" $( \zeta^a_{\ny k} + \eta^a_{\ny k})$, while the wave function associated with $\langle 0 | $ is an eigenfunction of the corresponding ``momentum"    $\partial / \partial (\zeta^a_{\ny k} + \eta^a_{\ny k})$. In similar fashion, as far as each  $\zeta^a_{\ny k} - \eta^a_{\ny k}$ variable is concerned, the wave function corresponding to $\langle 0 | $ is an eigenfunction of  the ``position" $( \zeta^a_{\ny k} - \eta^a_{\ny k}) $ while the wave function of $ | 0 \rangle $ is an eigenfunction of the corresponding ``momentum" $\partial / \partial ( \zeta^a_{\ny k} - \eta^a_{\ny k})$.  Therefore, the asymptotic  spaces of the bra and ket wave functions ---associated with the  creation/annihilation operators (42)--- have the typical features of the spaces ${\cal G}_-$ and ${\cal G}_+$ discussed in the previous sections. For this reason one can say that,  when  the operators $ u_{\pm }^a(\ny {k } ) $ and $v_{\pm}^a(\ny {k} )$ are expressed in the Schroedinger representation,  $\alpha_{\pm} (\ny k)$ and $\beta_{\pm }(\ny k)$  take essentially the form of the creation/annihilation operators in the representation (5), with the $*$-involution given in equation (43).  

To sum up, it turns out that the creation/annihilation operators that canonically define the Fock space ${\cal F}_2$ of the Chern-Simons theory in the Landau gauge  are related  to the  creation/annihilation operators which enter the mode decomposition  of the field operators  ---satisfying the reality condition (36)--- by means of the generalized Bogoliubov transformations   (42).

Equivalently, it turns out that  
the representation of the creation/annihilation operators, that canonically define the Hilbert space of the Chern-Simons theory in the Landau gauge, cannot be a $*$-representation of the fields operator algebra with respect to  the $*$-involution which is defined by the reality condition (36) for the bosonic fields  $A^a_\mu(x)$ and $B^a(x)$. This indicates that in the topological quantum field theories of the Chern-Simons type the structure of the asymptotic  state spaces  is  different from the vacuum  structure of conventional non-topological field theories. 

\vskip 0.4 truecm

\noindent {\sect 4.3. Minkowski signature  \blk }  When the gauge-fixing metric $\eta_{\mu \nu}$ has Minkowski signature, the time evolution is generated by an hermitian hamiltonian operator and, for this reason,  one can find partial similarities between the Chern-Simons equations of motion and the  standard elementary particles  dynamics.   So in this section let us consider  the case in which 
$$
\eta_{\mu \nu} dx^\mu d x^\nu = - (dx^1)^2 - (dx^2)^2 + (dt)^2 \; . 
\eqno(46)
$$ 
With the choice (46) of the gauge-fixing metric, the vacuum expectation values  of the Wilson line operators are not expressed in terms of conventional link invariants.  The reason is that the topology  which is induced by the ``distance" (46) in $\hbox {\matdop R}^3$ is not equivalent to the standard topology which is employed in knot theory.  For instance, the Feynman propagator for the $A^a_\mu(x)$ fields  in the Landau gauge ---computed with the metric (46)---  is no more given by the Gauss linking number expression that one obtains with euclidean signature.   However,   for the purposes of the present paper, the interest in the metric (46) does not concern the expectation values of the Wilson line operators; only the vacuum structure of the theory, in connection with the  representation of the fields operator algebra, will be considered. 

The field decomposition (31)-(34) together with the reality conditions (35) and (36) remain valid. The time evolution of $\{ u_{\pm}^a(\ny {k}, t)\} $ and $\{ v_{\pm}^a(\ny {k}, t)\}$  ---which is uniquely fixed by the free equations of motion---  now takes the form  
$$\eqalign { 
u_-^a (\ny {k} , t ) &= u_-^a (\ny {k} ) \cos (k t)  + v_-^{a  } ( \ny {k}  )  \sin ( k t ) \; , \;  
 v_-^a (\ny {k} , t ) = v_-^a (\ny {k}  )  \cos (k t)  - u_-^{a  } ( \ny {k}  )  \sin ( k t ) \; , \cr  
  u_+^{a} (\ny {k} , t ) &= u_+^{a} (\ny {k}  )  \cos (k t)  + v_+^a ( \ny {k}  )  \sin ( k t ) \; , \;  v_+^{a} (\ny {k} , t ) = v_+^{a}(\ny {k}  )  \cos (k t)  - u_+^a ( \ny {k}  )  \sin ( k t )  \; , \cr }
 \eqno(47)
$$
where the creation/annihilation operators  $\{ u_{\pm}^a(\ny {k})\} $ and $\{ v_{\pm}^a(\ny {k})\}$ satisfy the commutation relations (41) and $\left [ u_-^a(\ny {k}) \right ]^* = u_+^a(\ny {k}) $, $\left [ v_-^a(\ny {k}) \right ]^* = v_+^a(\ny {k}) $.  

The ``vacuum" vector $| \omega \rangle $ of the Fock space  ${\cal F}_1$, which is  canonically associated with $\{ u_{\pm}^a(\ny {k}, t)\} $ and $\{ v_{\pm}^a(\ny {k}, t)\}$,   corresponds to a stationary state but this state is not stable. Indeed, let $H_0$ denote the free hamiltonian and let $E_0$ represent  the energy of $| \omega \rangle $, $H_0 | \omega \rangle = E_0 | \omega \rangle$;  the state vector  
 $$\eqalign { 
 \left ( u^a_+ (\ny {k} , t ) + i v_+^{a} (\ny {k} , t ) \right )^N | \omega \rangle &= e^{ i t H_0}  \left ( u^a_+ (\ny {k}  ) + i v_+^{a} (\ny {k} ) \right )^N e^{-itH_0} | \omega \rangle \cr  & = e^{-iN k t}  \left ( u^a_+ (\ny {k}  ) + i v_+^{a} (\ny {k} ) \right )^N | \omega \rangle \cr } 
 \eqno(48)
 $$
has energy $E_N (\ny k ) $ given by 
$$
E_N (\ny k ) = E_0 - N k \; . 
\eqno(49)
$$ 
By varying $N$,  $E_N (\ny k ) $ can assume arbitrarily large negative values, $\lim_{N \rightarrow \infty}   E_N (\ny k ) = - \infty $. Note that the combinations $u^a_+ (\ny {k} , t ) - i v_+^{a} (\ny {k} , t ) $ would give positive energies, so that redefining $H _0\rightarrow - H_0$ would not solve the problem.  Therefore in the framework of the standard $*$-representations of the field operator algebra,  the energy spectrum  is not  bounded below and consequently this field theory model ---with the vacuum structure given by ${\cal F}_1$--- cannot describe a real   (stable) quantum mechanical system.    

On the other hand, let us consider the following creation/annihilation operators 
$$\eqalign { 
& \alpha_-^a (\ny {k } ) =  \mezzo \left [ u_-^a(\ny {k } ) + i v_-^{a} ( \ny {k } )  + u_+^a(- \ny {k } ) + i v_+^{a} (- \ny {k } )   \right ] \; , \cr
& \alpha_+^a (\ny {k } ) =  \mezzo \left [ u_+^{a} (\ny {k } ) - i v_+^{a} ( \ny {k } )  - u_-^{a}(- \ny {k } ) + i v_-^{a} (- \ny {k } )  \right ] \; , \cr
& \beta_-^a (\ny {k } ) = \mezzo \left [ u_-^a(\ny {k } ) + i v_-^{a} ( \ny {k } )  - u_+^a( - \ny {k } ) - i v_+^{a} (- \ny {k } )  \right ] \; , 
\cr 
& \beta_+^a (\ny {k } ) = \mezzo \left [ u_+^{a} (\ny {k } ) - i v_+^{a} ( \ny {k } )  + u_-^{a}( - \ny {k } ) - i v_-^{a} (- \ny {k } ) \right ]  \; ,  
\cr}
\eqno(50)
  $$
  with nonvanishing commutators  $\left [ \alpha_-^a(\ny {k } ) , \alpha_+^{b}( \ny {p}  ) \right ] = \delta^{ab} \delta (\ny {k} -  \ny {p}  ) = \left  [ \beta_-^a(\ny {k} ) , \beta_+^{b} ( \ny {p } ) \right ] $,   that diagonalize the free time evolution (47)
 $$
  \alpha_{\pm}^a (\ny {k } , t ) =  \alpha_{\pm}^a (\ny {k } ) e^{ \pm i k t } \quad , \quad  \beta_{\pm}^a (\ny {k } , t ) =  \beta_{\pm}^a (\ny {k } ) e^{ \pm i k t } \; . 
  \eqno(51)
 $$
  The $*$-involution acts as  
$$\eqalign { 
&\left [ \alpha^a_- (\ny {k}) \right ]^*  =   \beta^a_+( \ny{k})  \quad , \quad 
\left [ \alpha^a_+ (\ny {k}) \right ]^* =  \beta^a_-( \ny{k}) \; , \cr 
&\left [ \beta^a_- (\ny {k}) \right ]^* =  \alpha^a_+( \ny{k})  \quad , \quad  \left [ \beta^a_+ (\ny {k}) \right ]^*  =  \alpha^a_-( \ny{k}) \; . \cr } 
\eqno(52)      
$$
Let ${\cal F}_2$ be the Fock space which is canonically associated with $\alpha_{\pm} (\ny k)$ and $\beta_{\pm }(\ny k)$; the representation of the operator algebra defined in ${\cal F}_2$ is not a $*$-representation. Equations (51) show that the energy spectrum that is defined with respect to the Hilbert space structure associated with  the  Fock space ${\cal F}_2$ is bounded below; this means that ---as far as the energy stability is concerned--- the vacuum $| 0 \rangle \in  {\cal F}_2$ and  the annihilation/creation operators (50) describe a stable quantum mechanical system. 

 The   ``vacuum" vector $|0 \rangle \in {\cal F}_2 $ is defined by the relations $\alpha_{-} (\ny k) | 0 \rangle =0 $ and  $\beta_{-} (\ny k) | 0 \rangle =0 $ which are equivalent to  the conditions $\left [ u_-^a(\ny {k } ) + i v^{a}_- ( \ny {k } )  \right ] | 0 \rangle =0$ and $\left [ u^{a}_+ (\ny {k } ) + i v^{a}_+ ( \ny {k } )  \right ] | 0 \rangle =0$. If the annihilation and creation operators $u^a_-(\ny {k } ) $ and $ u^{a}_+ (\ny {k } )$ are represented  by  $ ( \zeta^a_{\ny k} +  \der / \der \zeta^a_{\ny k} ) / \sqrt 2$ and $ ( \zeta^a_{\ny k} -  \der / \der \zeta^a_{\ny k} ) / \sqrt 2$ and $v^a_-( \ny {k } ) $ and $ v^{a}_+ ( \ny {k } )$ are represented by $ ( \eta^a_{\ny k} +  \der / \der \eta^a_{\ny k} ) / \sqrt 2$ and $ ( \eta^a_{\ny k} -  \der / \der \eta^a_{\ny k} ) / \sqrt 2$, the vector $| 0 \rangle $ satisfies 
$$
\left ( \zeta^a_{\ny k} + i \eta^a_{\ny k} \right ) | 0 \rangle = 0 \quad , \quad \left ( \der / \der  \zeta^a_{\ny k} + i \der / \der \eta^a_{\ny k} \right ) | 0 \rangle = 0 \; . 
\eqno(53)
$$
On the other hand, conditions $\langle 0 | \alpha_+^a (\ny {k } ) =0 $ and $\langle 0 | \beta_+^a (\ny {k } ) =0  $ imply 
$$ 
\langle 0 |  \left ( \zeta^a_{\ny k} - i \eta^a_{\ny k} \right )  = 0 \quad , \quad \langle 0 | 
\left ( \der / \der  \zeta^a_{\ny k} - i  \der / \der \eta^a_{\ny k} \right )  = 0 \; . 
\eqno(54)
$$
The wave function representing the ket vector  $ | 0 \rangle $ must be an eigenfunction of  the ``position"  $(\zeta^a_{\ny k} + i \eta^a_{\ny k})$ and of the ``momentum" $\partial / \partial (\zeta^a_{\ny k} - i \eta^a_{\ny k}) $, whereas the wave function of the bra vector  $\langle 0 | $ must be an eigenfunction of the ``momentum"  $\partial / \partial (\zeta^a_{\ny k} + i \eta^a_{\ny k}) $ and of the ``position" $(\zeta^a_{\ny k} - i \eta^a_{\ny k})$. Again, the wave functions of   the asymptotic ket and bra state spaces are mutually related ---with respect to the appropriate variables--- in the same way as the functions belonging to ${\cal G}_-$ and ${\cal G}_+$ are. 

\vskip 0.4 truecm

\noindent {\sect 4.4. Semiclassical limit  \blk }  Traces of the peculiar vacuum structure of the topological quantum field theories of the Chern-Simons type  can also be found in the time evolution of quantum states in the  semiclassical  ---or large quantum numbers--- limit. The action  of the field operators $ A^a_\mu (x)$ on a given quantum state $| \chi \rangle  $ in the large quantum numbers ($LQN$) limit of quantum mechanics [9], 
$$
 \lim_{LQN} \,  A^a_\mu (x) | \chi , t \rangle \simeq {\cal A}_\mu^a (x) | \chi , t \rangle \; , 
\eqno(55)
$$
 defines a classical  field ${\cal A}_\mu^a (x) $ which satisfies the equations of motion (in the semiclassical approximation).   
In the $t \rightarrow \mp \infty $ limit,  ${\cal A}_\mu^a (x) $ determines the asymptotic wave functions   $\Psi_{in}$ and $\Phi_{out}$ that represent the ket $| \chi , t = - \infty \rangle $ and the bra $ \sigma | \chi , t = + \infty \rangle $ respectively.  The time evolution of ${\cal A}_\mu^a (x) $ describes  a semiclassical transition from $\Psi_{in}$ to $\Phi_{out} $. Since in the Chern-Simons theory canonically quantized in the Landau gauge the representation of the operator algebra is not a $*$-representation, the field configurations  on which $\Psi_{in}$ depends correspond to the proper values of ``conjugated" variables
with respect to the field configurations on which $\Phi_{out} $ depends; thus  the transition $\Psi_{in} \rightarrow \Phi_{out}$ must be realized by means of a nontrivial time evolution in the field variables.  This phenomenon, that can easily be verified for free fields, has significant consequences in the case of interacting fields;   it implies that at the semiclassical level the quantum Chern-Simons  theory necessarily describes a set of active evolving fields.  This is also confirmed by $(2+1)$ gravity. 
 
 In the Chern-Simons formulation [10] of (2+1) gravity, the identification of the intrinsic ``time" parameter is based on the structure of the gauge group $ISO(2,1)$. The set of real vector fields ---associated with the algebra of the group $ISO(2,1)$---  contains the triad components $e^a_\mu(x)$, with $a=1,2,3$, and the spin connection components $\omega^a_\mu(x)$. The gauge connection of the field theory can locally be represented by the 1-form $\cal W$ with value in the algebra of the gauge group ${\cal W} = e^a_\mu (x) P^a dx^\mu + \omega^a_\mu (x) J^a dx^\mu $, where $P^a$ are the generators of the spacetime translations (corresponding to the inhomogeneous component of $ISO(2,1)$) and $J^a $ are the generators of Lorentz group $SO(2,1)$.  

The gauge-fixed $ISO(2,1)$ lagrangian in the Landau gauge in the manifold $\hbox {\matdop R}^3$ is given by  [11] 
$$\eqalign { 
{\cal L}_g &= \mezzo \epsilon^{\mu \nu \rho} e^a_\mu \left ( \der_\nu \omega^a_\rho - \der_\rho \omega^a_\nu + \epsilon_{abc} \omega^b_\nu \omega_\rho^c \right ) + \der^\mu b^a e^a_\mu  + \der^\mu d^a \omega^a_\mu \cr
&+ \der^\mu {\overline \tau}^a \left ( D_\mu \tau \right )^a + \der^\mu \lambda^a \left ( D_\mu \xi \right ) + \epsilon_{abc} (\der^\mu \lambda^a ) e^b_\mu \tau^c \; . \cr } 
\eqno(56)
$$
According to the rules of the canonical quantization, the sets of fields $\{ e^a_\mu(x), d^a (x)\} $ and $\{ \omega^a_\mu(x), b^a (x)\} $, where $  b^a (x)$ and $d^a (x)$ denote the auxiliary fields, represent  conjugated variables.  This property persists  in the case of a  3-manifold of the type $ \Sigma \times \hbox {\matdop R}$ which admits a canonical quantization.  Essentially, the spin connection configurations can be understood as the classical values of the coordinates  or ``positions" of quantum mechanics,  whereas the triad  configurations represent  the  values  the corresponding ``momenta".
 Now, as far as the general solution of the classical equations of motion is concerned, the explicit functional dependence of the fields on the spacetime coordinates is not  known. However, the  3-manifolds whose geometry satisfies   the gravity equations are  known,  since  Mess has described [12] the general solution of (2+1) gravity,   the geometry  of the corresponding asymptotic states has been studied in Ref.[13] and the physical interpretation of the mathematical  construction of Mess has been developed by  Meusburger [14].    
  
Each Ricci flat  3-manifold $M$ with topology $ \Sigma_g \times \hbox {\matdop R}$, where  the closed two-dimensional spacelike surface $\Sigma_g $  has genus $g>1$,  describes  an expanding (or contracting) universe.  The presentation of these spacetimes  that has been produced by Mess is based on a purely geometric construction, so there are no meaningful   formulae to be reported here. For the purposes of the present paper, only  a brief description of the main features of the classical spacetime geometry needs to be given.   With respect to the gauge-invariant time parameter ---the cosmological time [15]---   which measures ``the length of time that events in $M$ have been in existence",   at the initial singularity  spacetime  degenerates into a codimension 2 real  tree [13] that identifies  a particular set of translation elements which belong to the inhomogeneous  component of $ISO(2,1)$, whereas the geometry of spacetime in the large time limit is determined by a subgroup $\Gamma $ of $SO(2,1)\subset ISO(2,1)$ that defines the hyperbolic structure of the space surface  $\Sigma_g \simeq \hbox{\matdop H} / \Gamma $, where {\matdop H} denotes the hyperbolic plane. This means that the wave functions which describe of the initial asymptotic states and the wave functions of the final asymptotic states  depend respectively on the proper values of conjugates variables.  

The spacetimes described by Mess give  the physical interpretation of the semiclassical approximation of the quantum states of  the  gauge $ISO(2,1)$ Chern-Simons field theory in the large quantum numbers limit.  All these spacetimes  have indeed a nontrivial cosmological evolution in which the semiclassical transition  connects asymptotic wave functions whose properties match the Chern-Simons vacuum structure described in the previous sections.    

To summarize, in the topological Chern-Simons quantum field theory  in the Landau gauge, the vacuum structure  gives origin to a representation of the fields  operator algebra which is not a $*$-re\-pre\-sen\-ta\-tion.  In the field theory perspective of the generalized Bogoliubov transformations (29),  the interplay between the different representations  for creation and annihilation operators  ---that have been described in the previous sections--- is relevant for the description of  the structure of the asymptotic wave functions; this structure  implies a nontrivial ``cosmological evolution" of the quantum states in the  semiclassical approximation.  

\vskip 0.7 truecm 

 \noindent {\bf Acknowledgments.}  I wish to thank M. Mintchev and  L.E. Picasso    for useful discussions. 
 
 \vskip 0.9 truecm

\noindent {\sect  Appendix A \blk }   
Let us consider the ${\rm C}^*$-algebra presentation [16] of quantum mechanics  and the main steps of the GNS construction [17,18], which clearly show how the algebraic relations among the operators control  the operator algebra representations. In order to simplify the exposition let us consider ---instead of a real ${\rm C}^*$-algebra---  the algebra ${\cal D}$   which is generated by the powers of $a_-$ and $a_+$ modulo the relation $[ a_- , a_+ ] = e $  where $e$ denotes the unit element of ${\cal D}$.  The star involution of ${\cal D}$ is fixed  by 
$ a_-^* = a_+ $ and $a_+^* = a_- $. 
 Each element  $d \in {\cal D}$ can  unambiguously be written in the canonical form 
$$ 
d = d_{00} \, e + \sum_{n\geq 1} d_{0n} \, a_-^n + \sum_{n \geq 1} d_{n0} \, a_+^n + \sum_{m \geq 1} \sum_{n \geq 1} d_{mn} \, a_+^m \, a_-^n \; ,  
\eqno(A1) 
$$
with complex coefficients $\{ d_{00}, d_{0n},  d_{n0}, d_{mn}\}$.  Let us consider the  pure state which is defined by the 
 positive linear functional  $\Omega : {\cal D} \rightarrow \hbox{\matdop C}$  that, when $d\in {\cal D}$ is expressed in the canonical form (A1), is given  by  
$$
\Omega : d \rightarrow \Omega (d) = d_{00}  \; . 
\eqno(A2)
$$
The left ideal $I_L$ of the elements  $d \in {\cal D}$ such that $\Omega (b d) =0 $ for any $b \in {\cal D}$ is given by   $I_L = {\cal D} a_-$. The linear space $V_-$ of the classes of ${\cal D} / I_L$ contains the elements  $ | ( d_{00} \, e + \sum_{n\geq 1} d_{n0} \, a_+^n  ) \rangle $. The canonical action of $\cal D$ on $V_-$ is given by $a | b \rangle = | ab \rangle $; therefore  
$$
 | ( d_{00} \, e + \sum_{n\geq 1} d_{n0} \, a_+^n  ) \rangle
=  ( d_{00} \, e +  \sum_{n\geq 1} d_{n0} \, a_+^n  ) | e \rangle \; ,  
\eqno(A3)
$$
where  the vector $| e \rangle \in V_- $ satisfies 
$$
a_- \, | e \rangle = 0 \; . 
\eqno(A4)
$$
Similarly, the right ideal $I_R$ of the elements  $d \in {\cal D}$ such that $\Omega ( d b) =0 $ for any $b \in {\cal D}$ is given by    $I_L = a_+ {\cal D} $. The linear space $V_+$ of the classes of ${\cal D} / I_R$ contains the elements $\langle  ( d_{00} \, e + \sum_{n\geq 1} d_{0n} \, a_-^n ) |$. The canonical action of $\cal D$ on $V_+ $ is given by $\langle b | a = \langle ba | $; then 
$$
\langle  ( d_{00} \, e + \sum_{n\geq 1} d_{0n} \, a_-^n ) | = \langle e |  ( d_{00} \, e +  \sum_{n\geq 1} d_{n0} \, a_-^n  )  \; , 
\eqno(A5)
$$
where  the vector $\langle e | \in V_+ $ satisfies
$$
\langle  e | \, a_+ = 0 \; . 
\eqno(A6)
$$
The linear space $V_-$ can be identified with the space of ket vectors of a quantum system, whereas $V_+$ corresponds to the space of bra vectors.  The pairing $ V_+ \times V_- \rightarrow \hbox{\matdop C}$  is defined by 
$$
\langle a | b \rangle = \Omega (ab) 
\; . 
\eqno(A7)
$$
A basis in  $V_-$  is given by the set of vectors $\{ | n \rangle =   (a_+^n / \sqrt {n!}  ) | e \rangle \} $ (with $n=0,1,2,...$);  
similarly, the vectors $\{ \langle m | =  \langle e | ( a_-^m / \sqrt { m!} ) \}$ form a basis in the space $V_+$, and $\langle m | n \rangle = \delta_{m n}$.

\vskip 2.5 truecm 

\noindent {\tmsm References} 

\vskip 0.7 truecm 

\item {[1]} V.~Bargmann, Commun. on Pure and Appl. Math. XIV (1961) 187. 

\medskip 

\item {[2]} B.~Bellazzini, M.~Mintchev and P.~Sorba, J. Math. Phys. 51 (2010) 032302. 

\medskip

\item{[3]}  N.N.~Bogoliubov, J. Phys. (USSR) 11 (1947) 23.

\medskip

\item{[4]} E.~Guadagnini, M.~Martellini and M.~Mintchev, Phys. Lett. B  227 (1989) 111. 

\medskip 

\item {[5]} E.~Guadagnini, J.  Phys. A: Math. Theor. 44 (2011) 415404,  ArXiv:1109.1384. 

\medskip

\item{[6]} E.~Witten, Commun. Math. Phys. 121 (1989) 351. 

\medskip 

\item {[7]} E.~Guadagnini, Phys. Lett.  B 251 (1990) 420. 

\medskip 

\item {[8]} E.~Guadagnini, Int. Journ. Mod. Phys.  A7 (1992) 877. 

\medskip

\item{[9]} L.~Landau et E.~Lifchitz, {\it M\'ecanique Quantique}, Editions MIR (Moscou, 1970). 

\medskip

\item{[10]} E.~Witten, Nucl. Phys. B 311 (1988) 46. 

\medskip

\item{[11]} E.~Guadagnini, N.~Maggiore and S.P.~Sorella, 
Phys. Lett. B 247 (1990) 543. 

\medskip 

\item{[12]} G. Mess, Geometriae Dedicata 126 (2007) 3-45, Preprint IHES/M/90/28 (1990). 

\medskip

\item{[13]} R.~Benedetti and E.~Guadagnini,  Nucl. Phys. B 613 (2001) 330-352, arXiv:gr-qc/0003055.  

\medskip 

\item{[14]} C.~Meusburger, Class. Quantum  Grav. 26 (2009) 055006,  arXiv:0811.4155; and   {\it Global lorentzian geometry from lightlike geodesic: what does an observer in (2+1)-gravity see?}, in Chern-Simons Gauge Theory: 20 Years After,  AMS/IP Studies in Advanced Mathematics vol. 50 (2011) 261, arXiv:1001.1842. 

\medskip 

\item{[15]} L. Andersson, G.J. Galloway and  R. Howard, Class. Quantum Grav. 15 (1998) 309.

\medskip 

\item {[16]}  R.~Haag and D.~Kastler, J. Math. Phys. 7 (1964) 848. 

\medskip

\item {[17]} I.M.~Gelfand and M.A.~Neumark, Isvertija Ser. Mat. 12 (1948) 445. 

\medskip 

\item {[18]} I.E.~Segal, Bull. Am. Math. Soc. 53 (1947) 73.

 \vfill\eject 
\end 
\bye